\documentclass[12pt,a4paper]{article}

\RequirePackage{amsmath}
\RequirePackage{mathrsfs}
\RequirePackage{amsfonts}
\RequirePackage{dsfont}
\RequirePackage{braket}
\RequirePackage{tabularx}
\RequirePackage{nicefrac}
\RequirePackage{graphicx}
\RequirePackage{booktabs}
\RequirePackage{colortbl}
\RequirePackage{xcolor}
\RequirePackage[symbol]{footmisc}

\def\AEF{A.E. Faraggi}

\def\IJMP#1#2#3{{\it Int.\ J.\ Mod.\ Phys.}\/ {\bf A#1} (#2) #3}

\def\EJP#1#2#3{{\it Eur.\ Phys.\ Jour.}\/ {\bf C#1} (#2) #3}

\def\JHEP#1#2#3{{\it JHEP}\/ {\bf #1} (#2) #3}
\def\NPB#1#2#3{{\it Nucl.\ Phys.}\/ {\bf B#1} (#2) #3}
\def\PLB#1#2#3{{\it Phys.\ Lett.}\/ {\bf B#1} (#2) #3}
\def\PRD#1#2#3{{\it Phys.\ Rev.}\/ {\bf D#1} (#2) #3}
\def\PRL#1#2#3{{\it Phys.\ Rev.\ Lett.}\/ {\bf #1} (#2) #3}
\def\PRT#1#2#3{{\it Phys.\ Rep.}\/ {\bf#1} (#2) #3}
\def\etal{{\it et al\/}}

\def\beq{\begin{equation}}
\def\eeq{\end{equation}}
\def\beqn{\begin{eqnarray}}
\def\eeqn{\end{eqnarray}}

\newcommand{\cc}[2]{c{#1\atopwithdelims[]#2}}
\newcommand{\CC}[2]{C{#1\atopwithdelims[]#2}}
\newcommand{\nn}{\nonumber}
\newcommand{\ba}{\begin{eqnarray}}
\newcommand{\ea}{\end{eqnarray}}

\begin{document}
\begin{titlepage}
\samepage{
\setcounter{page}{1}
\rightline{}
\rightline{June 2019}

\vfill
\begin{center}
{\Large \bf{String Phenomenology \\ \medskip
From a Worldsheet Perspective}}

\vspace{1cm}
\vfill

{\large Alon E. Faraggi$^{1}$\footnote{E-mail address: alon.faraggi@liv.ac.uk}
}\\

\vspace{1cm}

{\it $^{1}$ Dept.\ of Mathematical Sciences, University of Liverpool, Liverpool
L69 7ZL, UK\\}

\vspace{.025in}
\end{center}

\vfill
\begin{abstract}
\noindent
I argue that the ten dimensional non--supersymmetric tachyonic superstrings
may serve as good starting points for the construction of viable 
phenomenological vacua. Thus, enlarging the space of possible solutions that 
may address some of the outstanding problems in string phenomenology. 
A tachyon free six generation Standard--like Model is presented, 
which can be regarded as an orbifold of the $SO(16)\times E_8$ 
heterotic--string in ten dimensions. I propose that any
$(2,0)$ heterotic--string in four dimensions 
can be connected to a $(2,2)$ one via an orbifold or by interpolations 
and provide some evidence for this conjecture. It suggests that any 
Effective Field Theory (EFT) model that cannot be connected to a $(2,2)$ 
theory is necessarily in the swampland, and will simplify the 
analysis of the moduli spaces of $(2,0)$ string compactifications.

\end{abstract}

\smallskip}
\end{titlepage}

\section{Introduction}
\normalsize

String theory provides a viable framework to explore the unification 
of all the fundamental matter and interactions. 
While string theory produces a consistent theory of perturbative quantum 
gravity, it accommodates consistently the gauge and matter structures
of the subatomic regime, including the chirality property of the 
electroweak interactions. No other contemporary theory achieves this feat.
Yet it is expected that the string character of this basic theory is 
only exhibited at energy scales that are far removed from those 
accessible to present day experiments. In that respect one 
would have to rely on the effective, point--like, field theory
limit of the string construction in order to confront the theory
with experiment. For that purpose it is vital to be able to 
identify the smooth Effective Field Theory (EFT) limit corresponding to 
particular string theory vacua. To date this identification is only 
possible in limited cases \cite{dkl}, and entail mostly the analysis of 
various supergravity theories that are EFT 
limits of the corresponding string theories.
The picture is murky in both directions, as
for the most part one does not know whether an actual 
supergravity theory has an origin in a string construction. 
Furthermore, while supersymmetry is a beautiful theoretical construction,
it is not clear whether nature makes use of it. 
It is therefore important to explore alternatives from the
point of view of the worldsheet string theory.

To date investigation of non--supersymmetric string models
were primarily conducted in the context of the tachyon free 
$SO(16)\times SO(16)$ ten dimensional heterotic string theory 
\cite{dh, gv,  itoyama, kltclas, nonsusy, interpol, aafs}. 
This construction can be obtained as an orbifold of the ten
dimensional supersymmetric $E_8\times E_8$ heterotic--string, 
and can be connected to it by interpolations in a 
compactified dimension \cite{gv, itoyama}. It is well known that 
string theory gives rise to additional vacua in ten 
dimensions that are tachyonic \cite{dh, gv, kltclas}. 
However, the tachyonic modes may be projected out by Generalised 
GSO projections. 
Therefore, from the point of view of the worldsheet 
string theory, these string vacua may serve as
viable starting points for the construction of phenomenological 
string models and offer novel perspectives on some outstanding
issues in string phenomenology. Furthermore, they may 
reveal alternative symmetries to those provided by 
spacetime supersymmetry. An example is the Massive 
Boson--Fermion Degeneracy of \cite{msds}. 

In this paper I explore this possibility. For general reference I first 
examine the constructions of such models in ten dimensions. The discussion 
then reverts to the construction of phenomenological models in four 
dimensions that can be regarded as compactifications of the 
ten dimensional vacua. I present a six generation tachyon free model
with standard--like gauge group, as well as a three generation model
that does, however, contain two tachyonic states.
I discuss the reduction of the number of generations to three generations 
and the prospect of generating such tachyon free models. 

Additionally, the moduli spaces of (2,0) heterotic--string vacua is discussed. 
This class of compactifications is vast with little understanding of the
relation between the worldsheet constructions and their smooth
effective field theory
limits. I conjecture that all (2,0) heterotic--string vacua can be 
related to those with (2,2) worldsheet supersymmetry by an orbifold or by 
an interpolation and offer
some evidence for this conjecture, as well as some counter arguments. 
If the conjecture is correct it can serve as an enormous simplifying 
tool for the analysis of the moduli spaces of (2,0) string 
compactifications. In the very least it can serve as a classifying 
criteria between the vacua that can be regarded as descending 
from (2,2) vacua and those that do not.

\section{Ten dimensional vacua}\label{tendvacua}

We start our discussion with the $E_8\times E_8$ 
heterotic--string in ten dimensions. Its partition function is given by

\beq
 Z^+_{10d}= \frac{1}{{\tau_2}^4{(  \eta \overline \eta)}^8}
({V}_8-{S}_8)
\left( \overline O_{16} + \overline S_{16}\right)
\left( \overline O_{16} + \overline S_{16}
\right), 
\label{z10dplus}
\eeq
where the level--one $SO(2n)$ characters are given by
\beqn
O_{2n} &=& {\textstyle{1\over 2}} \left( {\vartheta_3^n \over \eta^n} +
{\vartheta_4^n \over \eta^n}\right) \,,
\nonumber \\
V_{2n} &=& {\textstyle{1\over 2}} \left( {\vartheta_3^n \over \eta^n} -
{\vartheta_4^n \over \eta^n}\right) \,,
\nonumber \\
S_{2n} &=& {\textstyle{1\over 2}} \left( {\vartheta_2^n \over \eta^n} +
i^{-n} {\vartheta_1^n \over \eta^n} \right) \,,
\nonumber \\
C_{2n} &=& {\textstyle{1\over 2}} \left( {\vartheta_2^n \over \eta^n} -
i^{-n} {\vartheta_1^n \over \eta^n} \right) \,
\label{thetacharacters}
\eeqn
In the following I omit the prefactor due to the 
uncompactified dimensions. The ten dimensional $SO(16)\times SO(16)$ 
heterotic--string is obtained by applying the orbifold projection
\beq
g = (-1)^{F+F_{z_1}+F_{z_2}}
\label{o16o16o}
\eeq
where $F$ is the spacetime fermion number, taking $S_8\rightarrow -S_8 $
and $F_{z_1,z_2}$ are the fermion numbers of the two $E_8$ factors, taking
$S_{16}^{1,2}\rightarrow -S_{16}^{1,2}$. The partition function of the 
$SO(16)\times SO(16)$ heterotic--string is therefore given by 
\ba
 Z_{10d}^-~=~  %\frac{1}{{\tau_2}^4{(  \eta \overline \eta)}^8}~~
&\left[ \right. & 
V_8 \left(\overline O_{16} \overline O_{16}+ 
           \overline S_{16} \overline S_{16}\right)
\cr
- && 
S_8 \left( \overline O_{16} \overline S_{16}+
            \overline S_{16} \overline O_{16}\right)
\cr
+ &&  
O_8 \left( \overline C_{16}  \overline V_{16} + 
           \overline V_{16}  \overline C_{16}\right) 
\cr
- && \left.
C_8 \left( \overline C_{16} \overline C_{16}  + 
           \overline V_{16} \overline V_{16} \right) \right]. 
\label{z10dminus}
\ea
Examining the partition function in eq. (\ref{z10dminus}) it is noted
that the would--be tachyonic term, 
$O_8 \left( \overline C_{16}  \overline V_{16} + 
           \overline V_{16}  \overline C_{16}\right) $
only produces massive physical states. Upon compactifications 
to lower dimensions, in general, tachyonic states will appear 
in the spectrum, but may be projected in special cases. 
In that respect, we may consider the ten dimensional tachyonic 
vacua and similarly project the tachyons in special cases. 

In the free fermion formulation \cite{fff} the vacua are specified
in terms of boundary condition basis vectors and one--loop Generalised
GSO (GGSO) phases. The $E_8\times E_8$ and $SO(16)\times SO(16)$ models are 
specified in terms of a common set of basis vectors 
\ba
v_1={\mathds{1}}&=&\{\psi^\mu,\
\chi^{1,\dots,6}| \overline{\eta}^{1,2,3},
\overline{\psi}^{1,\dots,5},\overline{\phi}^{1,\dots,8}\},\nonumber\\
v_{2}=z_1&=&\{\overline{\psi}^{1,\dots,5},
              \overline{\eta}^{1,2,3} \},\nonumber\\
v_{3}=z_2&=&\{\overline{\phi}^{1,\dots,8}\},
\label{tendbasisvectors}
\ea
where I adopted the conventional notation used in the phenomenological 
free fermionic constructions 
\cite{fsu5, slm, alr, lrs, su421, gkr, fknr, fkr, cfkr, acfkr, 
su62, frs, slmclass, lrsclass}. 
The basis vector $\mathds{1}$ is required by the consistency rules 
\cite{fff} and generates a model with an $SO(32)$ gauge
group from the Neveu-Schwarz (NS) sector. 
The spacetime supersymmetry generator 
is given by the combination 
\beq
S={\mathds{1}}+z_1+z_2 = \{{\psi^\mu},\chi^{1,\dots,6}\}. 
\label{tendsvector}
\eeq
The choice of GGSO phase $\CC{z_1}{z_2}=\pm1$ then selects between 
the $E_8\times E_8$ or $SO(16)\times SO(16)$ heterotic--string vacua 
in ten dimensions. The relation in eq. (\ref{tendsvector} dictates that 
in ten dimensions the breaking pattern $E_8\times E_8\rightarrow SO(16)\times
SO(16)$ is correlated with the breaking of spacetime supersymmetry. Eq. 
(\ref{tendsvector}) does not hold in lower dimensions.

To consider the tachyonic ten dimensional vacua we can start with 
the $E_8\times E_8$ partition function and apply the orbifold 
\beq
 g =  (-1)^{F+F_{z_1}        } \,, \label{torbifold}
\eeq
the resulting partition function is now given by
\beq
\left(V_8\overline O_{16} 
     -S_8\overline S_{16} 
     +O_8\overline V_{16} 
     -C_8\overline C_{16} \right)
\left(\overline O_{16} + \overline S_{16} \right),
\label{so16e8parfun}
\eeq
produces the partition function of the $SO(16)\times E_8$ 
non--supersymmetric and tachyonic heterotic--string vacuum. 
It is noted that the term $O_8\overline V_{16}\overline O_{16}$ 
in the partition function gives rise to a tachyonic 
state in the vectorial $16$ representation of $SO(16)$.
All of the non-supersymmmetric tachyonic string vacua in ten dimensions 
were classified in refs \cite{dh, kltclas}. It was further shown that
all the ten dimensional vacua can be connected by interpolations 
in lower dimensions or by orbifolds \cite{gv, itoyama}.

In the free fermion construction all the ten dimensional are 
specified in terms of boundary condition basis vectors and GGSO 
phases \cite{fff}. The $SO(16)\times E_8$ vacuum is generated by 
the basis vectors $\{{\mathds1}, z_1\}$ from eq. (\ref{tendbasisvectors}), 
irrespective of the choices of the GGSO phases. Other ten dimensional
vacua can similarly be generated by replacing the $z_1$ basis vectors
with $z_1=\{{\bar\phi}^{1,\cdots, 4}\}$ and additional similar $z_i$ 
basis vectors with utmost two overlapping periodic fermions. All 
these vacua are in principle connected by interpolations or orbifolds
along the lines of ref. \cite{gv}, and, in general, will contain
tachyons in their spectrum. Our interest here is rather in the possibility of 
constructing tachyon free phenomenological vacua, starting from the 
tachyonic ten dimensional vacua. The lesson to draw from the 
ten dimensional exercise is that these models can be constructed
by removing the ten dimensional vector 
$S={\mathds1}+z_1+z_2$ from the 
basis of the phenomenological four dimensional models.

\section{Lower dimensional constructions}\label{fourdmodels}

We can similarly consider compactifications of 
either of the models to lower dimensions, {\it e.g.} for the 
$E_8\times E_8$ heterotic--string 
\beq
{ Z}_+ = ( V_8 - S_8) \, \left( \sum_{m,n} \Lambda_{m,n}
\right)^{\otimes 6}\, \left(  \overline O _{16} + \overline S_{16} \right) 
\left( \overline O _{16} +  \overline S_{16} \right)\,,
\label{e8xe8partin4d}
\eeq
where for each circle,
\beq
p_{\rm L,R}^i = {m_i \over R_i} \pm {n_i R_i \over \alpha '} \,
\eeq
and
\beq
\Lambda_{m,n} = {q^{{\alpha ' \over 4} 
p_{\rm L}^2} \, \bar q ^{{\alpha ' \over 4} p_{\rm R}^2} \over |\eta|^2}\,.
\eeq
In the case of one compactified dimension, 
the $Z_+$ partition function is
\beq
{ Z}_+^{9d} = ( V_8 -  S_8) \,  \Lambda_{m,n} \,
		\left( \overline  O_{16} +  \overline S_{16} \right)
		\left(  \overline O_{16} +  \overline S_{16} \right)\,.
\label{zplusin9d}
\eeq
Applying the orbifold projection
\beqn
g &=& (-1)^{F_{z_1}+F_{z_2}}\delta \,, 
\label{deltaorbifold}
\eeqn
where 
$\delta x^9 = x_9 +\pi R$, 
in $Z_+^{9d}$ produces the $Z_-^{9d}$ partition function given by
\beqn
{ Z}_-^{9d} = (V_8 - S_8)~\left[ \right. 
& & \left. \Lambda_{2m,n} \,~~~~~~
\left(\overline O_{16}\overline O_{16}\, +
             \overline S_{16}\overline S_{16}\right) \right. \nonumber\\
&+& \left. \Lambda_{2m+1,n} \, ~~~
\left(\overline C_{16}\overline C_{16}~+
            \overline V_{16}\overline V_{16}\right)\right. \nonumber\\
 &+& \left. \Lambda_{2m,n+{1\over2}} \, ~~~
\left(\overline S_{16}\overline O_{16}~+
            \overline O_{16}\overline S_{16}\right)\right. \nonumber\\
 &+&  \left. \Lambda_{2m+1,n+{1\over2}} \, 
\left(\overline V_{16}\overline C_{16}~+
            \overline C_{16}\overline V_{16}\right)\right]~.\nonumber\\
\label{zminus9d}
\eeqn
The partition function of the free fermion model $\{1,S,z_1,z_2\}$, 
with $1+S+z_1+z_2=\{y^1 ,\omega^1~| ~\bar{y}^1,\bar{\omega}^1\}$, 
is given by
\beqn 
Z_{9d}&=&{1\over 2^4}
\left( \theta_3^4-\theta_4^4-\theta_2^4-\theta_1^4\right) 
\left\{
\left(|\theta_3|^2+|\theta_4|^2+|\theta_2|^2+|\theta_1|^2\right)
\left(\bar{\theta}_3^{16}+\bar{\theta}_4^{16} + \bar{\theta}_2^{16} 
+ \bar{\theta}_1^{16}\right)\right.
\cr
\cr
&&+\left[|\theta_3|^2+|\theta_4|^2+\CC{z_1}{z_2}
\left(|\theta_2|^2+|\theta_1|^2\right)\right]
\left[\bar{\theta}_3^8\bar{\theta}_4^8+
\bar{\theta}_4^8\bar{\theta}_3^8+\CC{z_1}{z_2}
\left(\bar{\theta}_2^8\bar{\theta}_1^8+
\bar{\theta}_1^8\bar{\theta}_2^8\right)\right]
\cr
\cr
&&+\left[|\theta_3|^2+\CC{z_1}{z_2}
\left(|\theta_4|^2+|\theta_1|^2\right)+|\theta_2|^2\right]
\left[\bar{\theta}_2^8\bar{\theta}_3^8+\bar{\theta}_3^8\bar{\theta}_2^8+
\CC{z_1}{z_2}\left(\bar{\theta}_4^8\bar{\theta}_1^8+
\bar{\theta}_1^8\bar{\theta}_4^8\right)\right]
\cr
\cr
&&\left.\left[\CC{z_1}{z_2}
\left(|\theta_3|^2+|\theta_1|^2\right)+\theta_4|^2+|\theta_3|^2\right]
\left[\bar{\theta}_2^8\bar{\theta}_4^8+\bar{\theta}_4^8\bar{\theta}_2^8+
\CC{z_1}{z_2}\left(\bar{\theta}_3^8\bar{\theta}_1^8+
\bar{\theta}_1^8\bar{\theta}_3^8\right)\right]\right\}.\nonumber\\
\ea
In terms of the $SO(2n)$ characters shown in eq. 
(\ref{thetacharacters}) we have
\beq
 Z_+=(V_8-S_8)\left(|O_2|^2+|V_2|^2+|S_2|^2+|C_2|^2\right)
\left( \overline O_{16} + \overline S_{16}\right) \left( \overline O_{16} +
\overline S_{16}\right) 
\eeq
and
\ba
 Z_-= (V_8-S_8)&\times& \left[
\left(|O_2|^2+|V_2|^2\right)
                 \left(\overline O_{16} \overline O_{16}+ 
                       \overline C_{16} \overline C_{16}\right)
\right. \cr\cr
&& + 
\left(|S_2|^2+|C_2|^2\right) 
                \left( \overline S_{16} \overline S_{16}+
                       \overline V_{16} \overline V_{16}\right) 
\cr
\cr
&& + 
\left(O_2\overline{V}_2+V_2\overline{O}_2\right)
                \left( \overline S_{16} \overline V_{16} + 
                       \overline V_{16} \overline S_{16}\right) \cr\cr
&&  + \left. 
\left(S_2\overline{C}_2+C_2\overline{S}_2\right)
                \left( \overline O_{16} \overline C_{16} + 
                       \overline C_{16} \overline O_{16} \right) \right] ,
\ea

where the orbifold operation is \cite{lhs}
\beqn
 a & = & (-1)^{F_L^{\rm int}+F_{\xi_1}} \,, \nonumber\\ 
 b & = & (-1)^{F_L^{\rm int}+F_{\xi_2}} \,, \label{9dspecial}
\eeqn
where $F_L^{\rm int}$ acts in the  $9^{\text{th}}$ dimension.
The observation here is that in lower dimensions we can 
couple the projection of the spinorial states from the sectors
$z_1$ and $z_2$ with an action in an internal dimension, thus 
breaking $E_8\times E_8\rightarrow SO(16)\times SO(16)$
without breaking supersymmetry. We can similarly 
consider the compactifications to four dimensions
on an $SO(12)$ lattice that corresponds to the 
enhanced lattice at the free fermionic point.
The two partition functions are given by
\beqn
{Z}^{4d}_- =  
{({V}_8-{S}_8)}
&\times&\left[~\left( |O_{12}|^2~+~|V_{12}|^2 ~\right) 
\left( \overline O_{16} \overline  O_{16}+  
\overline C_{16}  \overline C_{16}\right)\right.
\cr
&& + \left( |S_{12}|^2~~+|C_{12}|^2 ~\right) 
\left(  \overline S_{16}  \overline S_{16}+
 \overline V_{16}  \overline V_{16}\right)
\cr
&& + \left( O_{12} \overline  V_{12} + V_{12} \overline  O_{12} \right)
\left(  \overline S_{16}  \overline V_{16} +  
\overline V_{16}  \overline S_{16}\right)
\cr
&& + \left. \left( S_{12}  \overline C_{12} +C_{12}  \overline S_{12} \right)
\left(  \overline O_{16}  \overline C_{16} +  
\overline C_{16}  \overline O_{16} \right) \right] \,,
\label{zminus4d}
\eeqn
and
\beq
{Z}^{4d}_+=
{({V}_8-{S}_8)}
\left[|O_{12}|^2+|V_{12}|^2+
|S_{12}|^2+|C_{12}|^2\right]
\left( \overline  O_{16} + \overline S_{16}\right) 
\left(  \overline O_{16} +  \overline S_{16}
\right) \,, \label{zplus4d}
\eeq
and are connected by
the orbifold \cite{partitions}
\beq
{Z}_- = {Z}_+ / a \otimes b \,,
\label{zminusfromzplus}
\eeq
with
\beqn 
a &=& (-1)^{F_{\rm L}^{\rm int} + F_{z_1}} \,,
\nonumber \\
b &=& (-1)^{F_{\rm L}^{\rm int} + F_{z_2}} \,. \label{orbzpm}
\eeqn
Where $F_{\rm L}$ is the fermion number for the ``left''
component in the expression 
of the internal lattice {\it i.e.}, 
the nontrivial action of this operator
is $F_{\rm L} S_{12}=- S_{12}$ and 
   $F_{\rm L} C_{12}=- C_{12}$. 
The projection in (\ref{zminusfromzplus}) and (\ref{orbzpm}) 
is defined at the free fermionic point and can be generalised 
to an arbitrary point in the moduli space. The important point to
note is that all these supersymmetric and non--supersymmetric 
vacua can be interpolated by compactifications to lower dimensional 
vacua \cite{interpol}. 
Reversing the order of the projections we can consider them 
as compactifications of the non--supersymmetric $SO(16)\times SO(16)$
heterotic--string that are connected by interpolations to 
the supersymmetric vacua on the boundary of the moduli space. 
Similar constructions and interpolations can be implemented 
for the other ten dimensional string vacua. 

To construct phenomenological four dimensional vacua that correspond
to compactification of the ten dimensional tachyonic vacua, 
we can investigate the phenomenological free fermionic models. 
This class of heterotic--string models correspond to $Z_2\times Z_2$ 
orbifold of six dimensional toroidal lattices \cite{z2xz2}. 
From the analysis of the ten dimensional vacua we learn that 
the construction of the tachyonic ten dimensional vacua 
amounts to removing the vector combination $S$ from the 
allowed combination of basis vectors.
To construct a non--supersymmetric phenomenological four 
dimensional model, we can start with $SO(16)\times SO(16)$ ten 
dimensional model and explore the compactifications to four
dimensions on $Z_2\times Z_2$ orbifolds. 
This was pursued in ref. \cite{aafs}, 
and a general discussion of the tachyonic producing sectors 
was presented. In general, in addition to the tachyon 
that arise in the Neveu--Schwarz ${\vec 0}$ sector, 
the models contain numerous additional tachyon producing sectors. 
Those were classified in ref. \cite{aafs}. For specific choices of the
GGSO phases, phenomenological tachyon free models can be constructed 
\cite{aafs}.
The alternative is to explore compactifications of the 
ten dimensional tachyonic vacua. As discussed in section 
\ref{tendvacua} this amounts to removing the vector $S$
from the additive group in these constructions. 

Phenomenological free fermionic heterotic--string models were 
constructed by pursuing two methodologies. The first, which is 
referred to as NAHE--based models, was followed by using a 
common subset of boundary condition basis vectors, 
the so--called NAHE--set \cite{nahe}, that was first used
in the construction of the flipped $SU(5)$ (FSU5) 
heterotic--string model \cite{fsu5}
and subsequently employed in the construction of the 
Standard--like Models (SLM) \cite{slm}; 
Pati--Salam (PS) \cite{alr};
Left--Right Symmetric (LRS) \cite{lrs};
$SU(4)\times SU(2)\times U(1)$ (SU421) \cite{su421}; 
models. The NAHE--set is a common set of five basis vectors, 
$\{{\mathds1}, S, b_1, b_2, b_3\}$, where $S$ is the 
spacetime supersymmetry generator, discussed above, 
and $b_1$, $b_2$ and $b_3$ are the three twisted sectors 
of the $Z_2\times Z_2$ orbifold. The different phenomenological 
models are constructed by adding three or four additional 
boundary condition basis vectors to the NAHE--set 
\cite{fsu5,slm,alr,lrs,su421}. The second method 
entails a systematic
classification of toroidal $Z_2\times Z_2$ orbifolds for 
the different $SO(10)$ subgroups. The method was 
initially developed for the classification of vacua 
with an $SO(10)$ GUT group \cite{fknr,fkr} 
and subsequently employed for the classification of 
PS \cite{acfkr}; 
FSU5 \cite{frs}; 
SLM \cite{slmclass};
and LRS \cite{lrsclass} 
models. The method works with a fixed set of basis vectors and the
enumeration of the models is achieved by varying the
independent GGSO phases of the one--loop partition function, 
{\it i.e.} the set of basis vectors that generate the $SO(10)$ 
models is given by 

%\subsection{$SO(10)$ Models}

\begin{eqnarray}
v_1={\mathds{1}}&=&\{\psi^\mu,\
\chi^{1,\dots,6},y^{1,\dots,6}, \omega^{1,\dots,6}| \nonumber\\
& & ~~~\overline{y}^{1,\dots,6},\overline{\omega}^{1,\dots,6},
\overline{\eta}^{1,2,3},
\overline{\psi}^{1,\dots,5},\overline{\phi}^{1,\dots,8}\},\nonumber\\
v_2=S&=&\{{\psi^\mu},\chi^{1,\dots,6}\},\nonumber\\
v_{2+i}={e_i}&=&\{y^{i},\omega^{i}\; | \; \overline{y}^i,\overline{\omega}^i\},
\
i=1,\dots,6,\nonumber\\
v_{9}={b_1}&=&\{\chi^{34},\chi^{56},y^{34},y^{56}\; | \; \overline{y}^{34},
\overline{y}^{56},\overline{\eta}^1,\overline{\psi}^{1,\dots,5}\},\label{basis}\\
v_{10}={b_2}&=&\{\chi^{12},\chi^{56},y^{12},y^{56}\; | \; \overline{y}^{12},
\overline{y}^{56},\overline{\eta}^2,\overline{\psi}^{1,\dots,5}\},\nonumber\\
v_{11}=z_1&=&\{\overline{\phi}^{1,\dots,4}\},\nonumber\\
v_{12}=z_2&=&\{\overline{\phi}^{5,\dots,8}\},
\nonumber
\end{eqnarray}
where $i = 1,\ldots,6$ and the fermions which appear in the basis vectors have
periodic (Ramond) boundary conditions, whereas those not included have
antiperiodic (Neveu-Schwarz) boundary conditions.
Additional vectors are added to the set in (\ref{basis}) to 
generate the models with the various $SO(10)$ subgroups 
\cite{acfkr,frs,slmclass,lrsclass}. The GGSO phases $\CC{v_i}{v_{j}}$ 
with $i>j$ span the space of vacua, corresponding to $2^{n(n-1)/2}$
string models. 

The basis vector $S$ coincides with the vector combination in eq. 
(\ref{tendsvector}) and 
generates ${N} = 4$ spacetime supersymmetry, with $SO(44)$ 
gauge symmetry.
The $e_i$ vectors break the gauge symmetry to $SO(32)\times U(1)^6$ and
maintain the ${N}=4$ supersymmetry. These vectors correspond to all
possible internal symmetric shifts of the six internal bosonic coordinates.
The vectors $b_1$ and $b_2$ corresponds to ${Z}_2 \times {Z}_2$ 
orbifold twists. They break the spacetime supersymmetry
to ${N}=1$, and the gauge symmetry to $SO(10)\times U(1)^3 \times SO(16)$.
Addition of the basis vectors $z_1$ and $z_2$ breaks the hidden $SO(16)$
gauge group to $SO(8)\times SO(8)$.

The reduction of spacetime supersymmetry to $N=0$ can ensue by 
projecting the remaining supersymmetry from the $S$ basis vector. 
Setting $\cc{S}{v_i}=-\delta_{v_i}$
guarantees the existence of $N=1$ supersymmetry, 
and therefore the reduction to $N=0$ is obtained by 
relaxing this condition. The products $S\cdot e_i= 0$ and
$S\cdot z_i= 0$ entail that the $\{e_i, z_1, z_2\}$ basis vectors 
act as projectors on the $S$--sector. They can project all
the gravitinos from the $S$--sector, hence inducing the breaking from 
$N=4$ to $N=0$ spacetime supersymmetry.

Non--supersymmetric NAHE--based models as well as those
models generated by the set in eq. (\ref{basis}) can 
be represented as compactifications of the ten dimensional
$SO(16)\times SO(16)$ heterotic--string.
Similar to the ten dimensional
case the untwisted tachyonic state is projected out by the 
basis vector $S$. 
However, unlike the case of the ten dimensional model, 
the sectors that can produce 
additional tachyonic states proliferate. 
In the context of the three generation free fermionic models 
these sectors were classified in ref. \cite{aafs}. In general, 
the construction of realistic non--supersymmetric models without 
any tachyonic states is exceedingly hard. The reason is precisely 
due to the proliferation of tachyon producing sectors that arise 
due to the breaking of the string symmetries to smaller symmetries. 
In the free fermionic construction this is manifested by the larger
number of independent basis vectors that are required in the construction 
of the quasi--realistic three generation models. Examples of rare tachyon 
free cases can be found \cite{aafs}, and one can even search for such models
with suppressed cosmological constant \cite{nonsusy}.

As noted in the ten dimensional case compactifiations of the ten
dimensional tachyonic vacua amounts to removing the vector $S$
from the set of basis vectors, {\it e.g.} the set $\{{\mathds1}, z_1,z_2\}$
produces a non supersymmetric model with $SU(2)^6\times SO(12)\times 
E_8\times E_8$ or $SU(2)^6\times SO(12)\times SO(16)\times SO(16)$. 
In this case the untwisted tachyonic state in general reappear.
It is noted also that the left--moving vector bosons remain in 
the spectrum, and are projected out by the additional NAHE--set
basis vectors. We can start to explore such models by starting 
with a reduced NAHE--set that does not include the $S$--vector. 
This set is given by 
\beqn
 &&\begin{tabular}{c|c|ccc|c|ccc|c}
 ~ & $\psi^\mu$ & $\chi^{12}$ & $\chi^{34}$ & $\chi^{56}$ &
        $\bar{\psi}^{1,...,5} $ &
        $\bar{\eta}^1 $&
        $\bar{\eta}^2 $&
        $\bar{\eta}^3 $&
        $\bar{\phi}^{1,...,8} $ \\
\hline
\hline
      {\bf 1} &  1 & 1&1&1 & 1,...,1 & 1 & 1 & 1 & 1,...,1 \\
\hline
  ${b}_1$ &  1 & 1&0&0 & 1,...,1 & 1 & 0 & 0 & 0,...,0 \\
  ${b}_2$ &  1 & 0&1&0 & 1,...,1 & 0 & 1 & 0 & 0,...,0 \\
  ${b}_3$ &  1 & 0&0&1 & 1,...,1 & 0 & 0 & 1 & 0,...,0 \\
\end{tabular}
   \nonumber\\
   ~  &&  ~ \nonumber\\
   ~  &&  ~ \nonumber\\
     &&\begin{tabular}{c|cc|cc|cc}
 ~&      $y^{3,...,6}$  &
        $\bar{y}^{3,...,6}$  &
        $y^{1,2},\omega^{5,6}$  &
        $\bar{y}^{1,2},\bar{\omega}^{5,6}$  &
        $\omega^{1,...,4}$  &
        $\bar{\omega}^{1,...,4}$   \\
\hline
\hline
    {\bf 1} & 1,...,1 & 1,...,1 & 1,...,1 & 1,...,1 & 1,...,1 & 1,...,1 \\
\hline
${b}_1$ & 1,...,1 & 1,...,1 & 0,...,0 & 0,...,0 & 0,...,0 & 0,...,0 \\
${b}_2$ & 0,...,0 & 0,...,0 & 1,...,1 & 1,...,1 & 0,...,0 & 0,...,0 \\
${b}_3$ & 0,...,0 & 0,...,0 & 0,...,0 & 0,...,0 & 1,...,1 & 1,...,1 \\
\end{tabular}
\label{nahe}
\eeqn

The set of basis vectors in eq. (\ref{nahe}) produces a non supersymmetric 
model with 96 
multiplets in the $16$ spinorial representation of $SO(10)$. The four 
dimensional gauge symmetry is $SO(10)\times SO(6)^3\times E_8$. The model 
contain an untwisted tachyonic state in the vectorial 10 representation of 
$SO(10)$. This tachyonic state is entirely projected out in the 
FSU5 and SLM type models, but not in the PS or LRS models. Similar to the 
case of the non--supersymmetric models in ref. \cite{aafs}, whether or 
not a model contains tachyons is highly model dependent. The model 
defined by the set of basis vectors 
\begin{eqnarray}
1 &=&\{\psi^\mu,\
\chi^{1,\dots,6},y^{1,\dots,6}, \omega^{1,\dots,6}|
\bar{y}^{1,\dots,6},\bar{\omega}^{1,\dots,6},
\bar{\eta}^{1,2,3},
\bar{\psi}^{1,\dots,5},\bar{\phi}^{1,\dots,8}\},\nn\\
b_{1} &=& \{\psi^{\mu},\chi^{1,2},
y^{3,...,6}|\overline{y}^{3,...,6},\overline{\psi}^{1,...,5},\overline{\eta}^{1}\}
\nn\\
b_{2} &=& \{\psi^{\mu},  \chi^{3,4},
y^{1,2},\omega^{5,6}|\overline{y}^{1,2},\overline{\omega}^{5,6},
\overline{\psi}^{1,...,5},\overline{\eta}^{2}  \}\nn\\
b_{3} &=& \{\psi^{\mu}, \chi^{5,6}, \omega^{1,...,4}|
\overline{\omega}^{1,...,4}, \overline{\psi}^{1,...,5}, \overline{\eta}^{3}
\}\label{slmtachfree}\\
\alpha  &=& 
\{y^{1,...,6},\omega^{1,...,6}|\overline{\omega}^{1},
\overline{y}^{2},\overline{\omega}^{3},\overline{y}^{4,5},
\overline{\omega}^{6},\overline{\psi}^{1,2,3},\overline{\phi}^{1,...,4}\}\nn\\
\beta &=&\{y^{2},\omega^{2}, y^{4}, \omega^{4}
|\overline{y}^{1,...,4},\overline{\omega}^{5}, \overline{y}^{6},
\overline{\psi}^{1,2,3},\overline{\phi}^{1,...,4}\}\nn\\
\gamma &=&\{y^{1}, \omega^{1}, y^{5}, \omega^{5} | \overline{\omega}^{1,2},
\overline{y}^{3},\overline{\omega}^{4},\overline{y}^{5,6},
\overline{\psi}^{1,...,5}=\frac{1}{2}, 
\overline{\eta}^{1,2,3}=\frac{1}{2},
\overline{\phi}^{2,...,5}=\frac{1}{2}\}\nn
\end{eqnarray}
with the set of GGSO projection coefficients given by
\begin{center}
\beq
\bordermatrix{~ & 1 &b_{1}&b_{2}&b_{3}&\alpha&\beta&\gamma \cr
              1 &~~1&-1&-1&-1&~~1&\,\,\,\,1&\,\,\,\,i \cr
           b_{1} &-1&-1&-1&-1& -1&       -1&\,\,\,\,1 \cr
           b_{2} &-1&-1&-1&-1&~~1&       -1&\,\,\,\,1 \cr
           b_{3} &-1&-1&-1&-1& -1&\,\,\,\,1&\,\,\,\,1 \cr
              \alpha
&\,\,\,\,1&\,\,\,\,1&-1&\,\,\,\,1&\,\,\,\,1&\,\,\,\,1&\,\,\,\,1\cr
              \beta &\,\,\,\,1&-1&-1&-1&-1&-1& -1\cr
              \gamma &~~1&\,\,\,\,1&-1&\,\,\,\,1&-1&-1&\,\,\,\,1 \cr}.
\label{nonsphases}
\eeq
\end{center}
produces a tachyon free SLM model with 
\beqn
{\rm observable:}& & 
SU(3)_C\times  SU(2)_L\times U(1)_C\times U(1)_L\times U(1)^6
\label{oggroup}\\
{\rm hidden:}& &
 SU(5)_H\times SU(3)_H\times U(1)^2 \label{hggroups}
\eeqn
observable and hidden gauge groups, where the hidden sector gauge 
symmetry is generated by vector bosons that arise in the Neveu--Schwarz
sector and the sector $\zeta={\mathds1}+b_1+b_2+b_3$. 
The model contains six chiral 
generations in the spinorial $16$ representation of $SO(10)$, decomposed 
under the gauge group in eq. (\ref{oggroup}), from the sectors
$b_1$, $b_2$ and $b_3$. The doubling of the number of generations 
compared to the NAHE--based models occurs because of the removal 
of the $S$ projection, with the result that the chirality 
of the $\chi^{ij}$ worldsheet fermions in the sectors 
$b_1$, $b_2$ and $b_3$ is not fixed and consequently the number of 
generations is doubled. I discuss below how this can be remedied. 
The model contains three pairs of untwisted Higgs doublets 
$h_1, {\bar h}_1$, $h_2, {\bar h}_2$, $h_3, {\bar h}_3$,
that couple to the twisted states from the sectors 
$b_1$, $b_2$ and $b_3$ and produce a leading mass term for
the top quark mass. 
Additional electroweak Higgs doublet representation are obtained
from the sector $b_1+b_2+\alpha+\beta$. 
In that respect, the flavour structure in the model is similar 
to that of other NAHE--based Standard--like Models \cite{fh}.
The untwisted Neveu--Schwarz sector and the sectors
$\beta\pm\gamma$;
$\alpha\pm\gamma$;
$\alpha+\beta$;
$b_2+b_3+\beta\pm\gamma\oplus \zeta$;
$b_2+b_3+\alpha+2\gamma$;
$b_1+b_3+\alpha\pm\gamma\oplus \zeta$;
$b_1+b_3+\alpha+2\gamma$;
$b_1+b_2+\alpha+2\gamma$;
$b_1+b_2+\alpha+\beta$,
produce spacetime bosons, whereas the sectors
$b_1$;
$b_2$;
$b_3$;
$b_1+2\gamma \oplus \zeta$;
$b_2+2\gamma \oplus \zeta$;
$b_3+2\gamma \oplus \zeta$;
$b_2\pm\gamma$;
$b_1+b_2+b_3+2\gamma$,
produce spacetime fermions. Here the notation $\oplus \zeta$ denotes the 
states that transform under the hidden non--Abelian group factors, 
that are obtained from a given sector and the given sector $\oplus \zeta$.
The $U(1)_{1,2,3}$ symmetries are anomalous. Thus the model contains
one anomalous $U(1)$ combination that can be canceled by 
a generalised Green-Schwarz mechanism \cite{dsw}. The entire 
spectrum of the model will be presented elsewhere. 

As discussed above the model defined by eqs. (\ref{slmtachfree}, 
\ref{nonsphases}) gives rise to six chiral generation due to the 
removal of the $S$ projection on the states from the sectors 
$b_1$, $b_2$ and $b_3$. The consequence is that the chirality of the
worldsheet fermions $\chi^{ij}$ in these sectors is not fixed, 
hence doubling the number of generations compared to the NAHE--based 
three generation models. This situation can be remedied by including a
basis vector that mimics the projection of the $S$ vector, but without 
generating spacetime gravitinos, which is achieved by modifying the 
boundary conditions of the right--moving worldsheet fermions in the 
basis vector $S$. An example of a vector that achieves this feat 
is given by 
\beq
{\cal S} =  \{\psi^{\mu},\chi^{1,..,6}| \overline{\phi}^{1, 4,5,6}
\}
\label{slm3genMassiveS}
\eeq
with the choice of GGSO projection coefficients
\beq
\CC{\cal S}{\mathds1}= 
C{{\cal S}\atopwithdelims[]{b_1}}=
C{{\cal S}\atopwithdelims[]{b_2}}=
C{{\cal S}\atopwithdelims[]{b_3}}=
-\CC{\cal S}{\alpha}=  
-\CC{\cal S}{\beta}=  
-\CC{\alpha}{\cal S}=1 
\label{slm3genMassiveSphases}  
\eeq
the resulting model contains three generation of chiral 
fermions from the sectors $b_1$, $b_2$ and $b_3$. The vector
${\cal S}$ in eq. (\ref{slm3genMassiveS}) does not give rise to 
any massless gravitinos and therefore the model is non supersymmetric. 
Vector bosons contributing to the hidden sector gauge group arise 
again in the untwisted NS--sector and the $\zeta$-sector, 
generating an $SU(3)\times SU(2)\times U(1)^5$ hidden sector
gauge group, whereas the observable gauge group coincides with 
the one in eq, (\ref{oggroup}). The model does, however, 
contain two tachyonic states from the sector 
${\cal S}+b_1+b_2+b_3+\alpha+\beta+2\gamma$, that are 
neutral under the observable gauge symmetry and 
charged under the hidden sector gauge group.
A systematic search for similar tachyon free 
three generation models can be pursued by using  
the free fermionic classification method with or without a 
modified ${\cal S}$ sector and there is no a priori 
reason to assume that they do not exist. The situation 
in that respect is similar to the proliferation of gauge 
symmetry enhancing sectors in these models, 
but typically there exist configurations of free phases
in which all the enhancing vector bosons are projected out. 

\section{Connectedness of $(2,0)$ and $(2,2)$ string vacua}

In section \ref{fourdmodels} I argued that the ten dimensional 
tachyonic string vacua may serve as good starting points for the
construction of viable four dimensional string models. These are 
not the traditional string vacua that are explored in Effective 
Field Theory (EFT) studies of string compactifications, 
which are focused on the approximate supergravity 
limit of the supersymmetric string models.
As the basis vector $S$ is the generator of spacetime supersymmetry,
these EFT limits are those that would be characterised as effective
limits of string vacua that contain the basis vector $S$,
in the different ten dimensional string theories, {\it e.g.} 
in the Type II superstring and heterotic--string.
It is noted that the worldsheet perspective may afford
alternative starting points for the exploration of 
the phenomenological application of string theory. 

The heterotic--string is particularly appealing from the 
point of view of the Standard Model data, as it accommodates the 
embedding of the chiral matter states in the spinorial $16$ representation
of $SO(10)$. The supersymmetric string compactifications in four dimensions 
may have $(2,2)$ worldsheet supersymmetry or $(2,0)$, where the first case 
correspond to heterotic--string vacua with $E_6$ gauge symmetry in four 
dimensions, whereas the second case correspond to vacua in which 
the $E_6$ symmetry is broken to $SO(10)\times U(1)$ and its subgroups. 
While the moduli spaces of the $(2,2)$ string theories, and their
EFT limits, are fairly well 
understood \cite{dkl}, that is not the case for those with 
$(2,0)$ worldsheet supersymmetry. Understanding the moduli
spaces of $(2,0)$ string vacua and their EFT limits is an 
important problem in string phenomenology.
It is therefore of interest to explore whether string theory
can offer some guidance from a worldsheet perspective. 

It is known that the ten dimensional vacua are connected via
orbifolds or by interpolations in lower dimensions 
\cite{ginsparg, gv, itoyama}. 
The interpolation among string vacua was also studied
in the context of four dimensional phenomenological 
string vacua \cite{interpol}. 
It has further been proposed that the different
superstring theories can be seen to be
contained in the bosonic string \cite{cent}.
In this section I propose that all $(2,0)$ heterotic--string vacua 
can be connected to $(2,2)$ heterotic--string vacua via orbifolds
or via interpolations. I present some evidence for this conjecture 
that stems from the classification of fermionic $Z_2\times Z_2$ 
orbifolds and the observation of spinor--vector duality in the 
space of these compactifications. It should be noted that this 
claim is unexpected from the point of view of the effective field
theory description of string vacua. Indeed, some 
constructions have been presented that do not seem to have an
underlying $(2,2)$ structure \cite{bp}. 

We can again turn to the free fermionic models to seek guidance. 
We can consider the extended NAHE--set basis with the vectors
$\{{\mathds1}, S, b_1, b_2, b_3, z_1 \}$ \cite{xmap}.
As discussed above the subset 
$\{{\mathds1}, S, z_1= {\mathds1}+b_1+b_2+b_3, z_2 \}$
generates a model with $N=4$ spacetime supersymmetry with
$S0(12)\times SO(16)\times SO(16)$ or $SO(12)\times E_8\times E_8$ 
four dimensional gauge group, depending on the GGSO phase 
$\CC{z_1}{z_2}$, corresponding to the partition functions in 
eq. (\ref{zminus4d}) and (\ref{zplus4d}), respectively. 
Applying the $Z_2\times Z_2$ twists produces a model
with $N=1$ spacetime supersymmetry and
$SO(4)^3\times SO(10)\times U(1)^3$
or 
$SO(4)^3\times E_6\times U(1)^2$ 
gauge symmetries. 
The untwisted internal moduli space is identical in the two cases
and consists of three K\"ahler and three complex moduli \cite{moduli}.
In the fermionic 
language the exactly marginal operators corresponding to the 
moduli fields take the form of worldsheet Thirring interactions 
\cite{Thirring1987}. The untwisted moduli fields correspond to 
untwisted scalar fields that parametrise this moduli space \cite{moduli}.  
The vacuum with the enhanced $E_6$ symmetry has $(2,2)$ worldsheet symmetry, 
whereas in the vacuum with $SO(10)$ symmetry the right--moving 
$N=2$ worldsheet supersymmetry is broken. 
In the $E_6$ case the twisted sector produces 24 representations in the 
27 representation of $E_6$. In these models these states decompose under 
$E_6\rightarrow SO(10)\times U(1)$ in the following way. 
The spinorial $16$ representations are obtained from the sectors 
$b_1$, $b_2$ and $b_3$, whereas the vectorial $10$ representations 
of $SO(10)$ are obtained from the sectors $b_j+z_1$, $j=1,2,3$. 
In addition to the vectorial $10$ representations the sectors 
$b_j+z_1$, $j=1,2,3$ produce the 24 copies of: the 
$SO(10)$ singlets in the $27$ representation of $E_6$;
an additional singlet that correspond to the twisted moduli; 
and additional 8 $E_6$ real singlets, giving a total of 32 
real states. Correspondingly, in the $SO(10)$ models the $(2,2)$ 
worldsheet supersymmetry is broken. The breaking is induced by the 
same GGSO phase of the $N=4$ spacetime supersymmetric vacuum, 
namely $\CC{z_1}{z_2}=-1$. The sectors $b_j$ still produce the 24 copies
of the $16$ spinorial representation of $SO(10)$. However, the 
sectors $b_j+z_1$ now produce 24 copies in the 16 vectorial representation 
of the hidden $SO(16)$ gauge group. The total number of physical states
from the sectors $b_j\oplus b_j+z_1$ is therefore preserved and is 
again 32. However, the simple identification of the twisted moduli is 
obscured. The two models are, however, connected by the discrete map 
$\CC{z_1}{z_2}=+1\rightarrow -1$ in the fermionic worldsheet language. 
In the orbifold language the map is between two distinct Wilson
lines \cite{ffmt}, {\it i.e.} it is part of the $N=4$ moduli space.

The $Z_2\times Z_2$ on the $SO(12)$ lattice produces 24 fixed points 
\cite{xmap, z2xz2}. The $Z_2\times Z_2$ at a generic point in the moduli
space has 48 fixed. The number of 48 fixed points is reduced to 24 
by acting with a freely acting shift on the six dimensional 
torus. The freely acting shift that reproduces the $SO(12)$ 
lattice at the free fermionic point is a generalisation 
of the one given in eq. (\ref{orbzpm}), and involves a 
non--geometric asymmetric shift of both the momenta and winding 
modes \cite{partitions}. This asymmetric shift correspond to the 
fact that the $SO(12)$ lattice is realised with a non--trivial
antisymmetric $B$--tensor field, at the free fermionic point in the 
moduli space \cite{z2xz2,xmap}. 
The moduli space correspond to the $N=4$ moduli space, 
which is parametrised by the six dimensional metric $G$, the 
anti--symmetric tensor $B$, and the Wilson lines $W$. 

A general classification of the $Z_2\times Z_2$ orbifolds with $SO(10)$ 
GUT symmetry using the free fermion methodology
was performed in \cite{fknr, fkr}. Two relevant observation were made. 
The first is that the enumeration of the vacua with different 
numbers of generations only depends on the GGSO phases of the 
subset of basis vectors that preserve the $N=4$ spacetime 
supersymmetry. Hence, the enumeration only depends on the 
moduli fields of the $N=4$ toroidal compactifications.
At this level the moduli space is connected by continuous 
interpolations. 
The action of the $Z_2\times Z_2$ orbifolds, which breaks 
$N=4\rightarrow N=1$ spacetime supersymmetry, merely projects to 
different number of generations, but the information is 
predetermined by the data of the $N=4$ toroidal lattice.
The $Z_2\times Z_2$ orbifold action also projects some of 
the moduli fields. The transformations between the different
vacua at the $N=1$ level are therefore discrete, rather than continuous. 

The second observation in the fermionic classification of the 
$Z_2\times Z_2$ orbifolds is the existence of a global symmetry
in the space of $(2,0)$ string compactifications, under the exchange
of spinor and vector representation of the $SO(10)$ GUT group, dubbed 
spinor--vector duality \cite{fkr, cfkr, ffmt}.
This duality can be interpreted as a discrete 
remnant of the enhanced $E_6$ symmetry, just as $T$--duality 
is a discrete remnant of the enhanced symmetry at the 
self--dual point \cite{tduality}.
The vacuum at the $E_6$ enhanced symmetry point, which possesses $(2,2)$ 
worldsheet supersymmetry, is self--dual under the
spinor--vector duality. The spectral flow operator 
of the right--moving $N=2$ worldsheet supersymmetry is the operator
that mixes between the spinorial and vectorial $SO(10)$ states in the 
$27$ representation of $E_6$.
In the $SO(10)$ vacua, in which the right--moving
worldsheet supersymmetry is broken, the spectral flow 
operator induces the map between the dual vacua. 
Now, from the point of view of the fermionic or orbifold 
constructions, the order of the $N=4$ deformation eq. (\ref{orbzpm}), or 
the $Z_2\times Z_2$ orbifold, does not matter. 
Thus, the $(2,0)$ vacua can be interpreted as orbifold 
deformations of the $(2,2)$ vacua.
It is further noted that this picture generalises to 
string compactifications with interacting CFTs \cite{panos}. 
The existence of similar symmetries is expected to be 
a general property of $(2,0)$ heterotic--string vacua
with an $SO(10)$ GUT symmetry. 

String compactifications with $(2,0)$ worldsheet supersymmetry,
in general, do not 
need to possess an $SO(10)$ GUT symmetry. The right--moving gauge
symmetry may remain entirely unbroken; it can be broken to 
smaller subgroups; or it can be realised as a higher level 
Kac--Moody algebra \cite{higherKac}. 
In the case of the ten dimensional theories, it was argued in 
\cite{gv} that all the ten dimensional theories are indeed connected
by interpolations or orbifolds. In the same vein, we may hypothesise that the 
variety of four dimensional theories are similarly connected.
Given that a large class descend from the underlying $N=4$ toroidal space, 
there exist an uplift from the $N=1$ theory to the $N=4$ theory, 
which is the inverse of the modding out procedure of the 
breaking from $N=4$ to $N=1$. In the $N=4$ theory, the moduli space 
is continuous, so we expect that indeed all the $(2,0)$ can be 
connected by interpolations or by orbifolds to the $(2,2)$ theories.
In the very least, we see that some classes of $(2,0)$ compactifications, 
{\it e.g.} those with an $SO(10)$ GUT symmetry, can be seen to arise
as deformations of those with $(2,2)$ worldsheet supersymmetry, 
and investigation
of their moduli spaces can be facilitated by analysing this deformation. 
Furthermore, if we regard the $SO(10)$ embedding of the Standard Model
spectrum as phenomenologically desirable, these 
cases are the ones that may be physically relevant.
Given that the spinor--vector duality extends to worldsheet compactifications
with interacting CFTs \cite{panos}, gives reason to hypothesise that the
same structure extends, albeit in a more intricate way, to string 
vacua with interacting internal CTFs.

\section{Discussion and Conclusion}\label{conclusion}

The validation of the Standard Model as providing viable parametrisation
of all sub--atomic observable phenomena, reinforces the possibility that 
further insight into the Standard Model parameters can only be 
obtained by fusing it with gravity. Among contemporary quantum gravity 
approaches, string theory is unique because its internal consistency 
requirements mandates the existence of the matter and gauge structures that
are the bedrock of the Standard Model. By that string theory provide the 
arena to develop a phenomenological approach to quantum gravity. However, 
given that the string scale is far removed from experimentally accessible 
scales, it is likely that string theory will only provide some initial 
values for the Standard Model parameters, and their confrontation with 
experimental data will be performed by utilising effective field theory
methods. An example of this line of thought is the calculation of the 
top and bottom quarks Yukawa couplings and the resulting prediction
of the top quark mass \cite{topprediction}, which is obtained 
by evolving the string extracted parameters to the experimentally
accessible scale.  
Relating string vacua to their effective field theory smooth limit
is therefore an important problem in string phenomenology, which at 
present is only understood in limited cases \cite{dkl, stefan}.
Improving the understanding of the effective field theory limit
of string vacua is therefore an important problem in string 
phenomenology. 
To date relating string vacua to low energy observables relies exclusively 
on the effective supergravity limit. An important question is therefore
to explore to what extent is supersymmetry a necessary component in the 
construction of viable string models. 
Non--supersymmetric non--tachyonic vacua were constructed
in the past as compactifications of the 
$SO(16)\times SO(16)$ heterotic--string in ten dimensions.
However, worldsheet string theory may offer the alternative
of staring with a tachyonic ten dimensional vacuum and projecting the
tachyons with the GGSO projections. 
In this paper, I constructed one such six generation model with SLM 
gauge symmetry and discussed the reduction to three generations. 
Such models are particularly interesting from the point of view 
of the MSDS constructions that do not utilise the $S$ basis vector, 
which is common to the supersymmetric and $SO(16)\times SO(16)$ 
constructions. Whether an actual tachyon free three generation 
model can be constructed remains to be seen, but it is clear
that if it exists, it will have very special structure, rather 
than generic. Furthermore, the possibility to freeze all moduli,
aside from the dilaton, in fermionic $Z_2\times Z_2$ orbifolds \cite{cfmt},
offers the prospect of a such a model that is not connected
to a tachyonic point anywhere in the moduli space. 

Additionally, I proposed that from the worldsheet perspective 
all heterotic--string vacua with $(2,0)$ worldsheet 
supersymmetry can be connected to those with $(2,2)$ via orbifolds 
or interpolations. If correct, it will facilitate 
the understanding of the moduli spaces of 
$(2,0)$ heterotic--string compactifications. The evidence 
relies on the connectivity of the $N=4$ moduli space and the
existence of global symmetries, such as the spinor--vector
duality, in the space to $(2,0)$ heterotic--string compactifications. 
In the very 
least it can serve as a useful classification criteria between $(2,0)$
vacua that can, and those that cannot, be connected 
via orbifolds or interpolations to those with $(2,2)$ worldsheet 
supersymmetry.
An affirmative conclusion will support the suggestion 
\cite{corfu2010} that while string vacua are 
distinct from the point of view of the low energy field theory, 
they are equivalent from the string worldsheet point of view. 
From the worldsheet string perspective different string vacua
merely exchange massive and massless states. The preservation of 
the total number of massless states, distributed among the different
group factors, hints that the quantum gravity consistency requirements 
only care about the total number of massless degrees of freedom, 
rather than about the transformation properties under the 
low scale effective field theory gauge group.

\section*{Acknowledgments}

I would like to thank Costas Bachas, Doron Gepner and Dan Israel
for useful discussions. I would like to thank the 
Galileo Galilei Institute for Theoretical Physics and 
INFN for hospitality and partial support 
during the workshop ``String Theory from a worldsheet perspective''; 
and the CERN Theoretical Physics division for hospitality,
where part of this work was conducted.

\bigskip
%\newpage

\bibliographystyle{unsrt}

\end{document}